\documentclass[final]{svjour2}
\usepackage{graphicx}
\usepackage{rotating}
\usepackage{amssymb}
\usepackage{mathptmx}
\usepackage[numbers]{natbib}
\makeatletter
\journalname{Journal of Low Temperature Physics}

\bibpunct{}{}{,}{s}{}{,}

\begin{document}

\newcommand{\hdblarrow}{H\makebox[0.9ex][l]{$\downdownarrows$}-}
\title{Domain Size Distribution in Segregating Binary Superfluids}

\author{Hiromitsu Takeuchi}

\institute{Department of Physics, Osaka City University, 3-3-138 Sugimoto, Sumiyoshi-ku,\\ Osaka, 558-8585, Japan
\email{hirotake@sci.osaka-cu.ac.jp}}

\maketitle

\begin{abstract}
Domain size distribution in phase separating binary Bose--Einstein condensates is studied theoretically by numerically solving the Gross--Pitaevskii equations at zero temperature.
We show that the size distribution in the domain patterns arising from the dynamic instability obeys a power law in a scaling regime according to the dynamic scaling analysis based on the percolation theory.
The scaling behavior is kept during the relaxation development until the characteristic domain size becomes comparable to the linear size of the system, consistent with the dynamic scaling hypothesis of the phase-ordering kinetics.
Our numerical experiments indicate the existence of a different scaling regime in the size distribution function,
 which can be caused by the so-called coreless vortices.

\end{abstract}

\section{Introduction}

A homogeneous mixture of Bose--Einstein condensates (BECs) of two distinguishable bosons undergo phase separation due to the dynamic instability forming a characteristic domain pattern when the repulsive inter-component interaction is larger than a criterion \cite{2008Pethick}.
Phase separation of the two-component BECs is considered spontaneous symmetry breaking (SSB) when the system has a kind of symmetry, that is to say, if the particle mass, the intra-component interaction, and the concentration of a component in the initial state equal, respectively, those of the other.
An order parameter describing the phase separation is a real scalar field representing the difference between concentrations of the two condensates.
A domain wall, an interface between domains of the two components, forms a kink in the real scalar field as a topological defect.
 The dynamic instability triggered by a random seeds develops into a complicated net work of domain walls and the mean domain size grows with time in its relaxation dynamics by decreasing the total length of the walls according to the phase-ordering kinetics \cite{2014Hofmann,1994Bray}.

Recently, it has been shown that the domain patterns in the relaxation dynamics of the SSB transition of two-component BECs exhibit a scaling behavior of the percolation theory \cite{2015Takeuchi}.
The theory \cite{1994Stauffer} aims to describe how a connected element under consideration ``percolates'' through random systems.
A cluster of the element that percolates through the system is called a percolating cluster.
The system shows the universal critical behaviors described by the theory, independent of the microscopic details of systems, when the occupation rate $p$ of the element is close to the percolation threshold $p_c$ over which there appears a percolating cluster.
The role of clusters in the percolation theory is played by domains of a component in binary superfluids of two-component BECs.
It was revealed that the threshold value is $p_c\approx 0.5$ for the symmetric binary BECs and the largest domain and an infinite domain wall sandwiched between two percolating domains keep their fractal behavior during the phase separation process \cite{2015Takeuchi}.

In this work, the domain size distribution in two-component BECs in phase separation is investigated by numerically solving the coupled Gross--Pitaevskii (GP) equations.
The size distribution and its time development are explained theoretically based on the percolation theory and the phase-ordering kinetics.

\section{Formulation}
We consider quasi-two-dimensional two-component BECs in a uniform system at zero temperature.
This system is well described by the macroscopic wave functions $\psi_j(x,y,t)=\sqrt{n_j}e^{i\theta_j}~(j=1,2)$ of the $j$th component, which obey the coupled GP equations \cite{2008Pethick}
\begin{eqnarray}
i\hbar \partial_t \psi_j=-\frac{\hbar^2}{2m}\nabla^2\psi_j+\sum_{k=1,2} g_{jk}|\psi_k|^2\psi_j 
\label{GPeq}
\end{eqnarray}
with $m$ and $g_{jj}=g$ being the atomic mass and the intra-component coupling constant of the $j$th component, respectively.
The coupling parameters are assumed to satisfy the phase separation condition $g_{12}=g_{21}>g>0$ \cite{2008Pethick}.

The norm $N_j=\int {\rm d}x {\rm d}y n_j$ is conserved quantity during the time development and we set $N_1=N_2=N/2$.
By neglecting the thickness and curvature of domain walls in phase separating BECs,
 we obtains $n_j\approx n \equiv N/L^2$ in the bulk space, a place far from domain walls in phase-separated condensates.

For simplicity, we consider a case of strong segregation with $g_{12}/g=2$.
In this case, the thickness of a domain walls is of order the healing length $\sim\xi \equiv \hbar/\sqrt{gmn}$ in the bulk space.
The occupation rate $p$ of the $1$st ($2$nd) component is written by $p=N_1/N$ ($p=N_2/N$).
In our case of $p=0.5$, the dynamics of domain growth obeys the dynamic scaling law \cite{2014Hofmann} and the largest domain has a non-interger fractal dimension $D_S=91/48$ according to the percolation theory \cite{2015Takeuchi}.
Our discussion demonstrated here will be essentially applicable similarly to the phase separation dynamics in the spinor condensates, discussed in Refs. \cite{2013Kudo,2015Williamson}.

\section{The Initial Domain Pattern}
A uniform state $\psi_j=\sqrt{n/2}$ is the initial state of our numerical experiments.
This state is dynamically unstable leading to a spontaneous breaking of $Z_2$ symmetry with $n_d\equiv n_1-n_2\to n$ or $n_d \to -n$, triggered by some small fluctuations.
Numerically, a small white noise is added at $t=0$ to trigger the dynamic instability.
The dynamic instability is characterized by the length $l_0$ and time $t_0$.
The characteristic scales are described by the Bogoliubov spectrum of this systems \cite{2008Pethick};
$l_0$ and $t_0$ correspond respectively to the wave length and the inverse of the imaginary part of the frequency of the Bogoliubov mode with the largest imaginary part.
If the dynamic instability is triggered by a small random seed,
a domain pattern at $t=t_0$ has the characteristic length $l_0$.
For our case of $g_{12}/g=2$, one obtains
\begin{eqnarray}
l_0 &= & \pi \xi,
\label{l_0}\\
t_0 &=& \frac{2\pi m\xi^2}{\hbar} \ln\frac{n}{2\delta^2},
\label{tau_0}
\end{eqnarray}
where $\delta$ is the amplitude of the initial random seed.

The mean inter-wall distance $l(t)=L^2/R(t)$ is obtained with the total length $R$ of the domain walls, where $L$ is the linear size of the square system of $-L/2\leq x\leq L/2$ and $-L/2\leq y \leq L/2$.
The length $R$ is calculated by integrating the length of the sides between the square computational meshes with $n_d>0$ and $n_d<0$. The mesh size is $0.4\xi$.
The distance $l(t)$ increases from $l=0$ at $t=0$ to $l\sim l_0$ at $t=t_0$, consistent with the above estimation.
After that, $l(t)$ represents a power law in the scaling regime,
 where the domain growth obeys the dynamic scaling law.
The scaling behavior will continue until $l(t)$ becomes comparable to $L$.

\section{Domain Size Distribution in the Initial Domain Pattern}

We first analyze the size distribution of domains in the initial domain pattern with $l=l_0$.
We write the number of domains, whose area takes a value between $S$ and $S+dS$, divided by the system area $L^2$
as $\rho(S,l)dS$.
If our system is described by the percolation theory,
the distribution function $\rho_0(S)\equiv \rho(S,l_0)$ of the initial pattern obeys the power law 
\begin{eqnarray}
\rho_0(S)\propto S^{-\tau}
\label{Fisher}
\end{eqnarray}
with the Fisher exponent $\tau\approx 2$ according to the percolation theory in two-dimensional systems \cite{1994Stauffer}.

\begin{figure}
\begin{center}
\includegraphics[%
  width=0.7\linewidth,
  keepaspectratio]{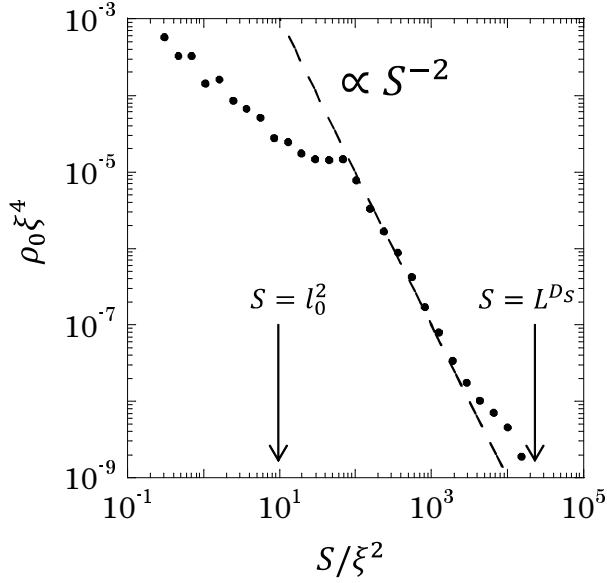}
\end{center}
\caption{
Domain size distribution $\rho_0(S)$ in the initial pattern with $l=l_0$.
A broken line represents the power law with the approximate value $\tau=2$ of the Fischer exponent.
The distribution $\rho_0(S)$ obeys the power law in the scaling regime $l_0^2 \ll S \ll L^{D_S}$ with the fractal dimension $D_S=2-\beta/\nu$ with the critical exponents $\beta=5/36$ and $\nu=4/3$ of two-dimensional percolation theory.}
\label{Nd_s}
\end{figure}

Figure \ref{Nd_s} shows the numerical result of the domain size distribution $\rho_0(S)$ for $L=204.8\xi$.
The plot fits the power law (\ref{Fisher}) with $\tau=2$ in a wide range of $S$.
The function $\rho_0(S)$ deviates from the power law behavior for $S/\xi^2 \sim 10^4$.
This is because, the scaling analysis is inapplicable statistically when the area is comparable to the system area $L^2\sim 10^4\times\xi^2$.
Particularly,
$\rho_0(S)$ must go down to zero around $S \sim L^{91/48}$
since the largest (percolating) domain has the fractal dimension $D_S=91/48$ according to two-dimensional percolation theory.
We have included the contribution of the percolation domain in the statistics. It is almost always that there exists a percolation domain, whose area is remarkably larger than that of other domains.
This contribution generates the overshoot on the tail of the distribution in Fig. \ref{Nd_s}. 

The lower bound of the scaling range should be $S\sim l_0^2$, which corresponds to the characteristic area of the domain patterns caused by the dynamic instability.
We may say that the length $l_0$ in the domain patterns plays the role of the smallest length scale in the percolation patterns on mosaics,
where the scaling behavior appear on the scale much larger than the smallest scale of the mosaic.
Therefore, the distribution $\rho_0(S)$ can reflect the scaling behavior (\ref{Fisher}) within the scaling regime $l_0^2 \ll S \ll L^{D_S}$.

\section{Dynamic Scaling Plot of the Domain Size Distribution}

According to the dynamic scaling law in the phase-ordering kinetics,
 the domain patterns at a later time are statistically similar to those at an earlier time by scaling with the characteristic length $l(t)$.
 We expect that the same holds for the domain size distributions $\rho(S,l)$.
 Rescaling the area $S$ as $\tilde{S}=Sl^{-2}$,
we may assume a universal form for the distribution function,
 \begin{eqnarray}
\rho(S,l)l(t)^{4}=c_S \tilde{S}^{-\tau}\equiv \tilde{\rho}(\tilde{S})
\label{Fisher_dynamic}
\end{eqnarray}
with a constant $c_S$ independent of $S$ and $l(t)$.

\begin{figure}
\begin{center}
\includegraphics[%
  width=0.7\linewidth,
  keepaspectratio]{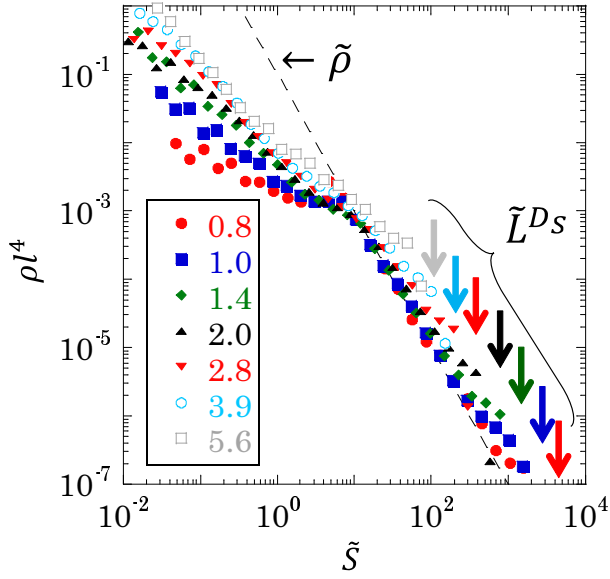}
\end{center}
\caption{Dynamic scaling plot of the domain size distribution $\rho(S,l)$ for $l(t)/l_0=$ 0.8, 1.0, 1.4, 2.0 ,2.8, 3.9, and 5.6 with the effective system size $\tilde{L}=L/l(t)=$81.5, 65.2, 46.6, 32.6, 23.3, 16.7, and 11.6, respectively. The broken line represents the universal function $\tilde{\rho}(\tilde{S})$ with $c_S=0.1$ and $\tau=2$.
The positions of $\tilde{S}=\tilde{L}^{D_S}$ for different values from $l(t)/l_0=0.8$ to $l(t)/l_0=5.6$ are represented by thick arrows from right to left. (Color figure online)  
}
\label{Nd_sFSS}
\end{figure}

Figure \ref{Nd_sFSS} shows the scaling plots of the distribution functions $\rho(S,l)$ at different times together with the universal function (\ref{Fisher_dynamic}) with $c_S=0.1$ and $\tau=2$.
 The scaling plots with $l(t)/l_0 \lesssim 2.8$ ($\tilde{L}\gtrsim 23.3$) coincide well with each other for $10\lesssim \tilde{S} \ll \tilde{L}^{D_S}$ consistent with Eq. (\ref{Fisher_dynamic}).
Note that the scaling regime for the initial domain pattern, $l_0^2 \ll S \ll L^{D_S}$, is reduced to $1 \ll \tilde{S} \ll \tilde{L}^{D_S}$ in the scaling plots.
 Here, $\tilde{L}=L/l(t)$ is the effective system size scaled with the shortest length of the ``mosaic'' pattern.
The width of the scaling regime in the plots increases with $\tilde{L}$,
 and then the regime almost disappears for $l(t)/l_0=$3.9 and 5.6 with $\tilde{L} \sim 10$.

It is interesting that the dynamic scaling plot seems to be asymptotic to a power law in the smaller regime $\tilde{S}\lesssim 1$.
The smaller regime describes the size distribution of domains whose area is smaller than the characteristic area $l^2$.
This regime can show the statistics of the smaller domains of a component which live in a ``sea'' of the other component, where the linear size of the ``sea'' is $l(t)$.
In this sense, this fact might indicate that the relaxation dynamics of smaller domain walls with size much smaller than $l(t)$ is also described by some scaling analysis. 
The number of such closed walls must be small at $t=t_0$ because the initial domain patterns do not have any structure whose length scale is much smaller than $l(t=t_0)=l_0$.
This argument gives us a reason why the number density for $\tilde{S}\lesssim 1$ increases with time monotonically.

The smaller closed walls shrink and subsequently vanish due to evaporation as is in the conventional system of binary liquid \cite{1994Bray}.
However, this is not the whole story in our system of binary superfluids.
A closed wall can be stabilized as a stable topological defect, called a coreless vortex,
 in which a component is trapped by the density depression caused by a quantized vortex in the other component.
A nature as a quantized vortex becomes important
 when the radius ($\sim \sqrt{S}$) of a coreless vortex is comparable to the thickness ($\sim \xi$) of the domain wall that surrounds the vortex core \cite{2013Hayashi}.

Such vortices can be nucleated from domain walls due to Kelvin--Helmholtz instability if a relative velocity between the two components is enough large across a domain wall \cite{2010Takeuchi}.
This hydrodynamic effect would be unimportant in the early stage ($t\lesssim t_0$)
 since the dynamic instability of the phase separation does not induce a relative motion due to momentum exchange between two condensates \cite{2011Ishino}.

The hydrodynamic effect can be important in the later development of the phase separation dynamics.
Then, a scaling regime unique to binary superfluids can exist during the later development in a range $\xi^2 \ll S \ll l^2$.
In fact, the scaling behavior in the range observed in our system is qualitatively different from those, $\tilde{\rho}(\tilde{S})\sim \tilde{S}^{1/2}$ for $\tilde{S} \ll 1$, predicted for conventional conserved fields at zero temperature \cite{2009Sicilia} although the system is not described by hydrodynamics.
It is an open problem how quantized vortices influence the scaling behavior of the size distribution.

\section{Conclusion}
We studied domain size distribution in phase separation dynamics induced by the dynamic instability in two-component BECs in quasi-two-dimensional systems.
The size distribution in the domain patterns after the characteristic time of the instability ($t=t_0$) obeys the scaling law (Eq. \ref{Nd_sFSS}) of two-dimensional percolation theory for domains in a scaling range $l^2 \ll S \ll L^{91/48}$.
The size distribution in the smaller regime $S \ll l^2$ approaches asymptotically to a scaling function after the number density of the smaller domains increase substantially.
 The scaling analysis in the smaller regime is an interesting future problem.

\begin{acknowledgements}
This work was partly supported by KAKENHI from JSPS (Grant No. 26870500).
\end{acknowledgements}

\pagebreak

\end{document}